\documentclass{PoS}

\rightline{\sffamily UTCCS-P-25, UTHEP-530, KEK-CP-186, HUPD-0610}\vspace*{-10mm}

\title{An estimate of the $\eta$ and $\eta^\prime$ meson masses in $N_f=2+1$ lattice QCD}

\ShortTitle{An estimate of the $\eta$ and $\eta^\prime$ meson masses in $N_f=2+1$ lattice QCD}

\author{CP-PACS and JLQCD Collaborations:}
\author{
   S.~Aoki${}^{b,g}$,   
   M.~Fukugita${}^f$,  
   K.-I.~Ishikawa${}^e$, 
   T.~Ishikawa${}^b$,
   N.~Ishizuka${}^{a,b}$, 
   Y.~Iwasaki${}^b$, 
   K.~Kanaya${}^b$, 
   Y.~Kuramashi${}^{a,b}$,
   M.~Okawa${}^e$, 
   Y.~Taniguchi${}^b$, 
   A.~Ukawa${}^{a,b}$,
   N.~Yamada${}^{c,d}$, 
   \speaker{T.~Yoshi\'e}${}^{\ a,b}$\thanks{Email: yoshie@ccs.tsukuba.ac.jp}\\
\\\llap{${}^a$} Center for Computational Sciences, University of Tsukuba,
                Tsukuba 305-8577, Japan\\
  \llap{${}^b$} Graduate School of Pure and Applied Sciences,
                University of Tsukuba, Tsukuba 305-8571, Japan\\
  \llap{${}^c$} High Energy Accelerator Research Organization (KEK),
                Tsukuba 305-0801, Japan\\
  \llap{${}^d$} School of High Energy Accelerator Science,
                The Graduate University for Advanced Studies (Sokendai),
                Tsukuba 305-0801, Japan\\
  \llap{${}^e$} Department of Physics, Hiroshima University,
                Higashi-Hiroshima 739-8526, Japan\\
  \llap{${}^f$} Institute for Cosmic Ray Research, University of Tokyo,
                Kashiwa 277-8582, Japan\\
  \llap{${}^g$} Riken BNL Research Center, Brookhaven National Laboratory,
                Upton, New York 11973, USA\\
}

\abstract{
Masses of the $\eta$ and $\eta^\prime$ mesons are estimated 
in $N_f=2+1$ lattice QCD 
with the non-perturbatively $O(a)$ improved Wilson quark action
and the Iwasaki RG-improved gluon action, using CP-PACS/JLQCD 
configurations on a $16^3\times 32$ lattice at $\beta=1.83$ 
(lattice spacing is 0.122 fm). 
We apply a stochastic noise estimator technique combined with 
smearing method to evaluate correlators among flavor $SU(2)$ singlet 
pseudoscalar operators and strange pseudoscalar operators 
for 10 combinations of up/down and strange quark masses. 
The correlator matrix is then diagonalized to identify signals 
for mass eigenstates. 
Masses of the ground state and the first excited state 
extrapolated to the physical point are $m_\eta=$ 0.545(16) GeV and
$m_{\eta^\prime}=$ 0.871(46) GeV, being close to the experimental values
of the $\eta$ and $\eta^\prime$ masses. 
}

\FullConference{XXIVth International Symposium on Lattice Field Theory\\
                July 23-28, 2006\\
                Tucson, Arizona, USA}

\begin{document}

\section{Introduction}
The U(1) problem~\cite{ref:weinberg} is an outstanding 
issue in hadron spectroscopy. 
The large mass of the $\eta^\prime$ relative to the $\pi$
is believed to be
related to QCD vacuum structure and anomaly of the axial current. 
In order to reproduce the $\eta^\prime$ mass 
from first principles calculations, 
a number of lattice QCD simulations have been carried out. 
Early attempts~\cite{ref:IIY,ref:kuramashi} were made in
quenched QCD. 
Recent studies~\cite{ref:sesam,ref:ukqcd,ref:cppacs} 
shifted to full QCD with two degenerate up and down ($u/d$) quarks.

In the real world the role of strange ($s$) quark is important since
the mass difference between the $s$ quark and $u/d$ quarks is 
the cause of mixing between the singlet and octet pseudoscalars, 
that leads to the physical $\eta$ and $\eta^\prime$ mesons.
In this article, we report on an attempt to estimate  
masses of the $\eta$ and $\eta^\prime$ in $N_f=2+1$ QCD,
in which degenerate $u/d$ quarks and an $s$ quark are treated
dynamically.
The mixing effect is taken into account by diagonalizing
a correlator matrix for operators coupled to both $\eta$ and 
$\eta^\prime$.
We use Wilson quark formulation.
An attempt to calculate the $\eta^\prime$ mass with staggered 
fermions is made in Ref.~\cite{ref:gregory}.

The calculation is carried out on a set of gauge configurations
generated previously for a study of the flavor non-singlet
hadron spectrum and light quark masses in $N_f=2+1$ 
QCD~\cite{ref:ishikawa}.
This report presents results for the $\eta$ and $\eta^\prime$ masses 
calculated for the coarsest of the three lattice spacings.  

A technical difficulty in the study of the $\eta$ and $\eta^\prime$ 
masses lies in the evaluation of the two-loop quark diagram contribution to 
the singlet part of propagators. 
Due to configuration fluctuations in the two-loop diagram, 
errors of the $\eta$ and $\eta^\prime$ propagators 
increase rapidly as time slice. 
On the other hand, we have to suppress contamination from
higher excited states.
Calculations have to be setup in a way that signals of 
the $\eta$ and $\eta^\prime$ appear at small time slices.
For this purpose, we combine a smearing method with a
stochastic noise estimator technique (SET), 
as carried out in Ref.~\cite{ref:cppacs}.

The organization of this article is as follows. 
Numerical calculations including our procedure to 
calculate the two-loop diagram are described 
in Sec.~\ref{sec:Calculation}.
In Sec.~\ref{sec:Analysis} we give an account of 
analysis procedure and the $\eta$ and $\eta^\prime$ masses
at the physical point.
Conclusions are given in sec.~\ref{sec:Conclusion}. 
Data presented here should be regarded as preliminary
since detailed analysis is still ongoing.

\section{Numerical Calculations}\label{sec:Calculation}
\subsection{Configuration}
We use a set of $N_f=2+1$ full QCD configurations~\cite{ref:ishikawa}
generated with the Iwasaki RG-improved gauge action and 
the non-perturbatively $O(a)$ improved Wilson fermion action 
at $\beta=1.83$ on a $16^3\times 32$ lattice. 
Polynomial hybrid Monte Carlo simulations have been carried out 
for 10 quark mass combinations;
we choose five hopping parameters $K_{ud}=$ 0.13655,
0.13710, 0.13760, 0.13800 and 0.13825 for the $u/d$ quarks,
and two $K_{s}=$ 0.13710 and 0.13760 for the $s$ quark. 
The pseudoscalar to vector mass ratio ranges from 0.78 to 0.61.
The lattice spacing is $a=$ 0.122 fm which is determined from 
experimental values of the $\pi$, $\rho$ and $K$ meson masses.

The trajectory length is 7000 -- 8600 for each mass combination.
Measurements are carried out at every 10 trajectories. Errors
are estimated by the jack-knife method with a bin size of 10 
configurations (100 trajectories).

\subsection{Two-loop quark diagram} 
When one determines the $\eta$ and $\eta^\prime$ masses
from propagators around a time slice, contamination from 
higher excited states has to be small at that time slice. 
For this condition to be satisfied, it is desirable that
contamination is small also for octet pseudoscalars.
In Ref.~\cite{ref:ishikawa}, we adopted an exponential-like smearing
function
\begin{equation}
  K({\vec n} - {\vec m}) = A \exp ( - B \vert{\vec n} - {\vec m}\vert ) 
     \ \ \mbox{\ for\ }\  {\vec n} \ne {\vec m}, \ \ \ \ K({\vec 0}) = 1, 
     \label{eq:kernel} 
\end{equation}
with $A=1.2$ and $B=0.1$, and find that effective masses of
octets obtained with doubly smeared source and unsmeared (point) 
sink operators reach an approximate plateau at a small time slice; 
the mass estimated at $t\approx 4$ is only a few percent
different from accurate value of the ground state mass.
With this observation in mind, we construct the two-loop diagram
using a doubly smeared loop diagram at the source with 
the kernel (\ref{eq:kernel}) and an unsmeared loop diagram at the sink, 
and combine it with the corresponding one-loop diagram.

The two-loop diagram is estimated by a combined method of SET 
and smearing applied to configurations fixed to Coulomb gauge.
We prepare a $U(1)$ random number $\exp(i\theta({\vec n},t))$ 
for each site $({\vec n},t)$ of the whole lattice and solve a quark 
propagator $q({\vec n},t)$ for the random source. 
The propagator is then smeared twice at a source time slice $t_{\rm src}$, 
multiplied by the inverse of the random number and 
summed over sites on the time slice,
\begin{equation}
  \xi(t_{\rm src}) = \sum_{{\vec l},{\vec m},{\vec n}} 
         \exp(-i\theta({\vec l},t_{\rm src})) 
         K({\vec l}-{\vec m})K({\vec m}-{\vec n})q({\vec n},t_{\rm src}).
\end{equation}
The loop diagram $\zeta(t_{\rm snk})$ at a sink time slice $t_{\rm snk}$ 
is calculated similarly without smearing. 
The two-loop diagram is estimated by averaging the product of the two loops
over 20 stochastic noise sets generated for each color and spin 
at a time,
\begin{equation}
 G_{\rm 2-loop}(t) = \frac{1}{L^3T} \sum_{t=t_{\rm snk}-t_{\rm src}}
                     \langle
                     \mbox{Tr}(\gamma_5 \zeta(t_{\rm snk}) )
                     \mbox{Tr}(\gamma_5 \xi(t_{\rm src}) )
                     \rangle_{\rm noise}.
\end{equation}

\begin{figure}
\begin{center}
\includegraphics[scale=0.62]{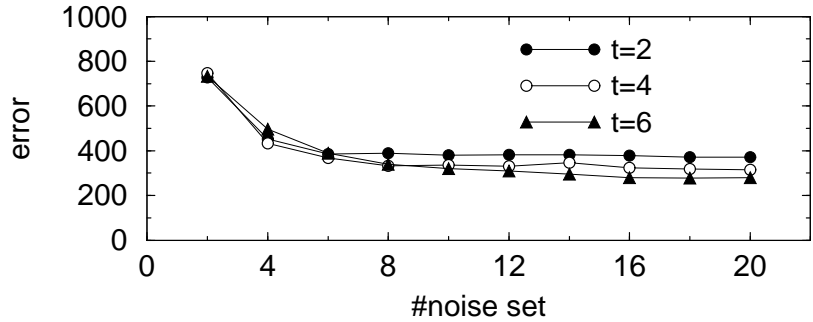}
\caption{Error of the two-loop diagram versus number of noise sets
for $(K_s=0.13710, K_{ud}=0.13825)$.
The vertical scale is arbitrary.}
\label{fig:errrnd}
\end{center}
\end{figure}

In any stochastic noise methods, error consists of stochastic
noise and configuration fluctuations. In our case, the error of 
the two-loop diagram decreases rapidly with a number of noise sets 
and reaches a plateau for a range of time slices we use
for later analysis. See Fig.~\ref{fig:errrnd}.
This means that our final error obtained with full 20 noise sets 
is dominated by configuration fluctuations.

\section{Analysis and Results}\label{sec:Analysis}
\subsection{$\eta$ and $\eta^\prime$ masses at simulation points}
In order to derive mases of the $\eta$ and $\eta^\prime$,
we consider flavor $SU(2)$ singlet and strange 
pseudoscalar operators
\begin{equation}
 \eta_n = (\bar u \gamma_5 u + \bar d \gamma_5 d)/\sqrt{2},\ \ 
 \eta_s = (\bar s \gamma_5 s),
\end{equation} 
and calculate the $2\times 2$ correlators
\begin{equation}
G(t) = 
\left(
  \begin{array}{cc}
    \eta^P_n(t)\eta^S_n(0) & \eta^P_n(t)\eta^S_s(0) \\
    \eta^P_s(t)\eta^S_n(0) & \eta^P_s(t)\eta^S_s(0) 
  \end{array}
\right),
\end{equation}
where superscripts $S$ and $P$ denote smeared and point operators.
If the correlator matrix is dominated by the contribution of 
the ground state ($\eta$) and the first excited state ($\eta^\prime$)
at a time slice $t$, we obtain a relation 
\begin{equation}
  (C G(t) G^{-1}(t_0) C^{-1})_{ij} \approx \delta_{ij} \exp(-m_i(t-t_0)) 
  \ \ \ \ \ \  i,j=\eta,\eta^\prime, \label{eq:propdecay}
\end{equation}
where the matrix $C$ diagonalizes the basis operators $\eta^P_k$ ($k=n,s$) to
give the physical ones $O_i$ ($i=\eta,\eta^\prime$),
$O_i = \sum_{k=n,s} C_{ik} \eta^P_k$, 
and $t_0$ is a reference time slice.
The matrix $C$ is independent of time slice. 
Selecting a representative time slice $t_D$, we estimate the matrix 
$C$ by diagonalizing the matrix $G(t_D)G^{-1}(t_0)$.
We then calculate the ``diagonalized'' propagator 
\begin{equation}
  D_{ij}(t) = (C G(t) G^{-1}(t_0) C^{-1})_{ij}. \label{eq:diag}
\end{equation}
Note that $D_{ij}(t_0)=\delta_{ij}$ and 
$D_{ij}(t_D)$=0 ($i\ne j$) by definitions.

\begin{figure}[t]
\begin{minipage}{75mm}
\begin{center}
\includegraphics[scale=0.51]{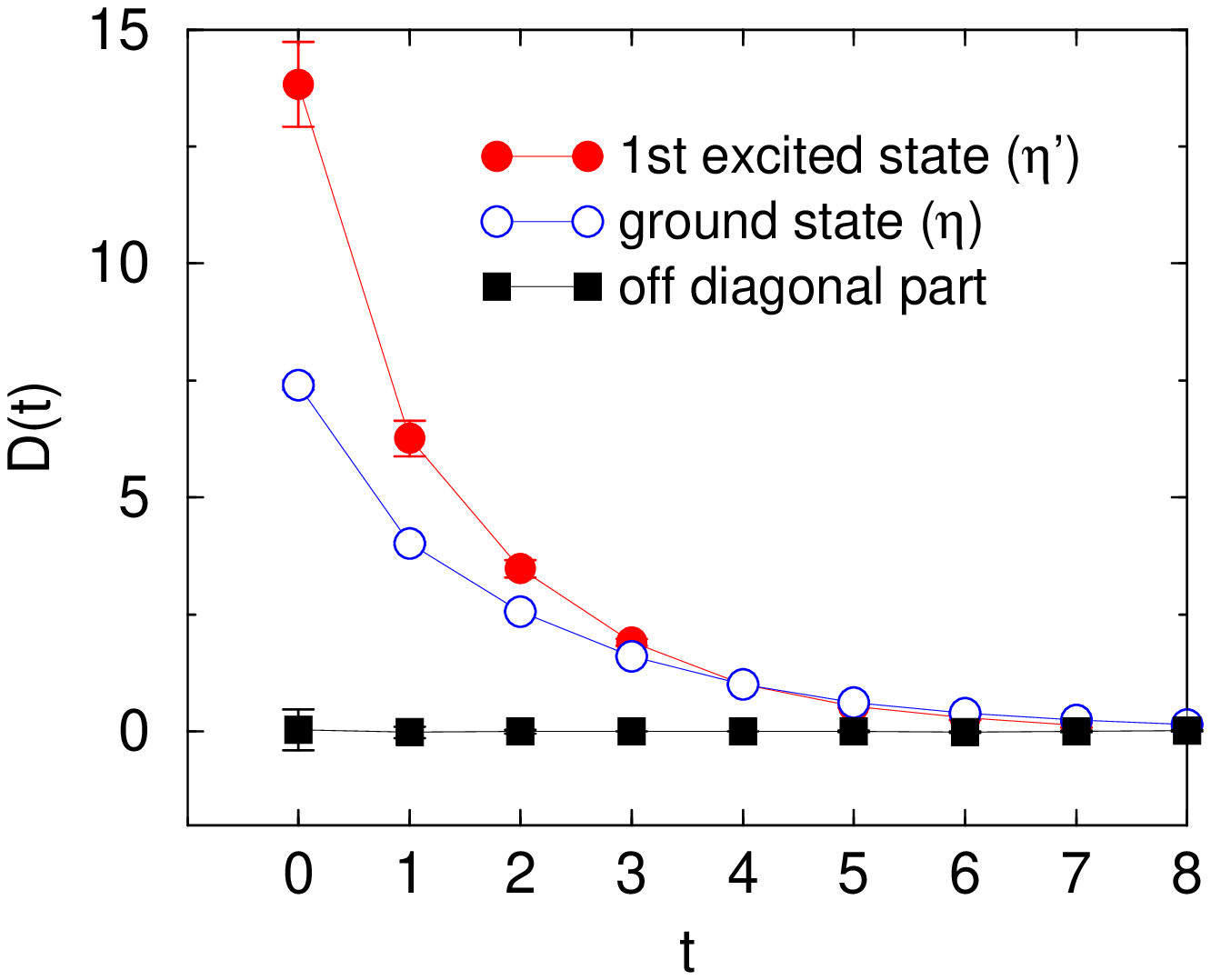}
\caption{Diagonalized propagator 
for $(K_s=0.13710, K_{ud}=0.13825)$. For off-diagonal
part, $D_{12}(t)$ in eq.~(\protect\ref{eq:diag}) are plotted.}
\label{fig:diagprop}
\end{center}
\end{minipage}\hspace{4mm}
\begin{minipage}{75mm}
\begin{center}
\includegraphics[scale=0.51]{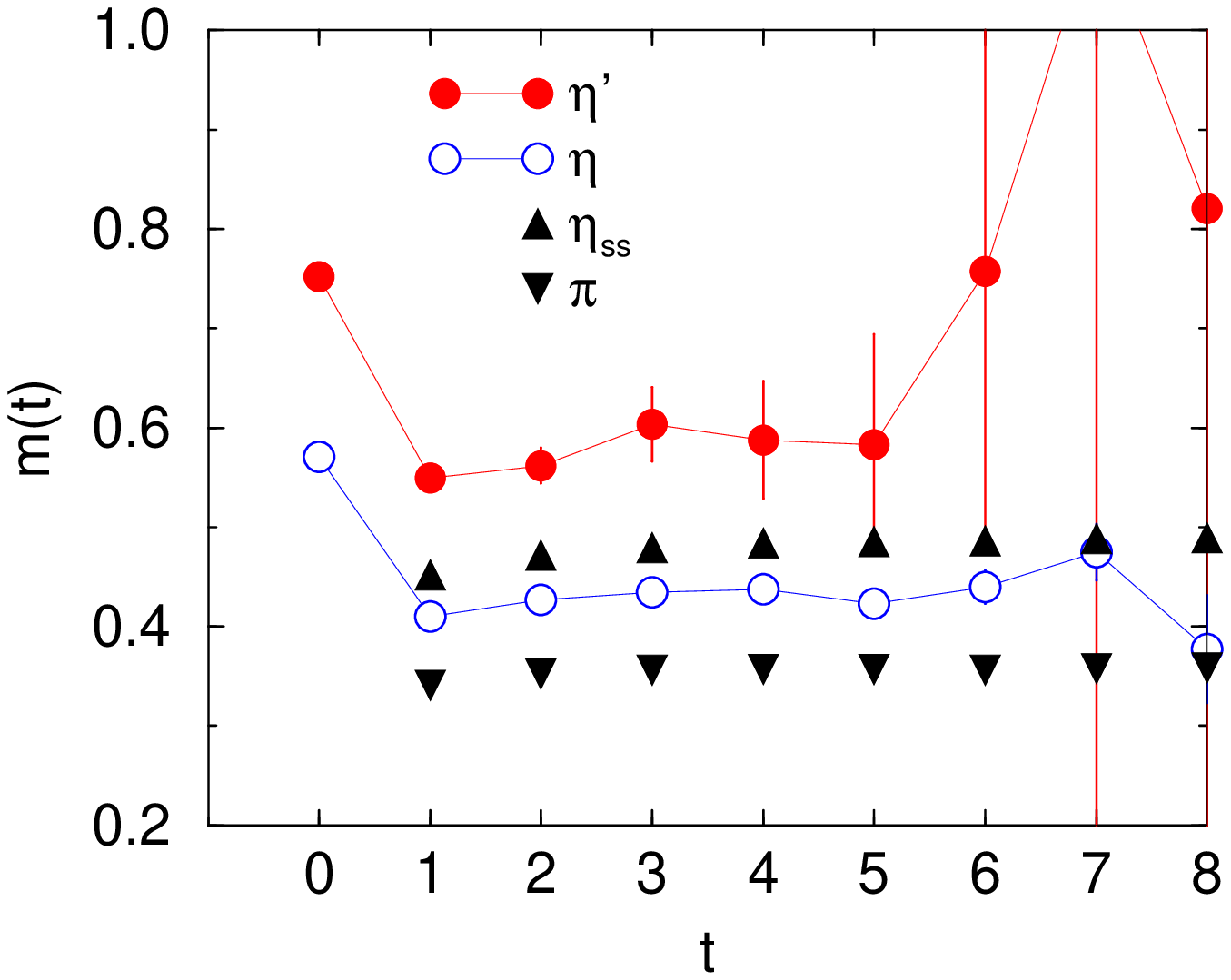}
\caption{Effective masses of the $\eta$ and $\eta^\prime$
for $(K_s=0.13710, K_{ud}=0.13825)$.
Those for $\pi$ and $\eta_{ss}$ are overlaid.}
\label{fig:effmass}
\end{center}
\end{minipage}
\end{figure}

Figure~\ref{fig:diagprop} shows the diagonalized propagator 
obtained with $t_0=4$ and $t_D=3$ for the mass combination of
the heaviest $s$-quark and the lightest $u/d$-quarks.
Diagonal parts (signals for $\eta$ and $\eta^\prime$) decay 
exponentially as indicated by effective masses plotted 
in Fig.~\ref{fig:effmass}, while off-diagonal parts are
negligible for all $t$. 
In other words, the relation (\ref{eq:propdecay}) is realized.
This result suggests that the $\eta$ and $\eta^\prime$ 
contribution dominates our correlator matrix at small time slice,
e.g., already at $t\approx 2$.

In Fig.~\ref{fig:effmass}, we overlay effective masses
of an octet pseudoscalar $\bar u \gamma_5 d$ ($\pi$) and 
those of $\bar s \gamma_5 s$ determined by ignoring 
the two-loop diagram contribution (``$\eta_{ss}$''). 
The mass of the $\eta^\prime$ is much larger than
masses of the $\pi$ and $\eta_{ss}$, which demonstrates 
the importance of the two-loop diagram contribution for 
the large mass of the $\eta^\prime$ 

\begin{figure}
\begin{center}
\includegraphics[scale=0.84]{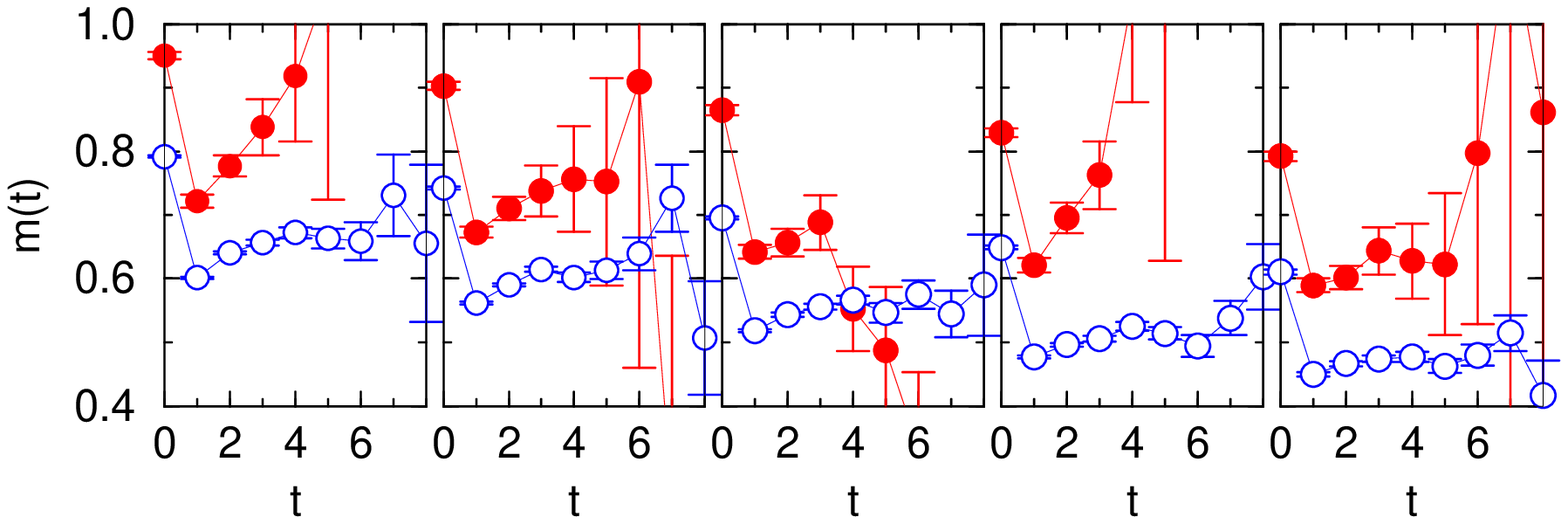}
\includegraphics[scale=0.84]{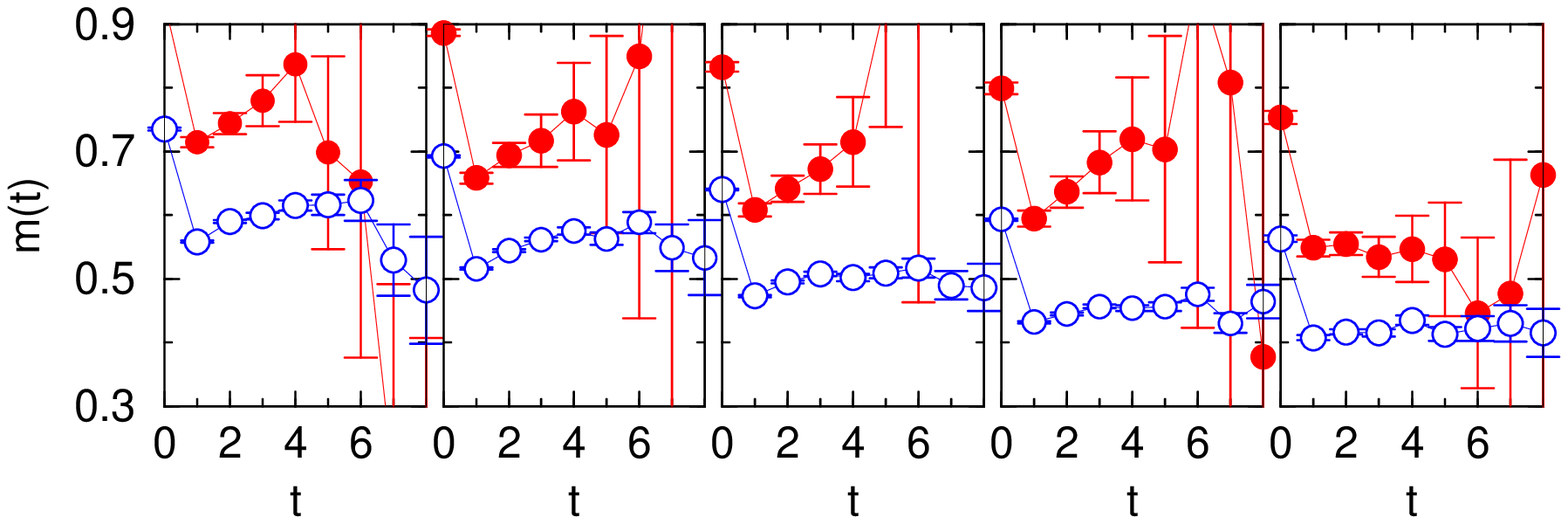}
\caption{Effective masses of the $\eta$ (open symbols) and 
$\eta^\prime$ (filled symbols) for 
$K_s=$0.13710 (top panels) and 0.13760 (bottom panels).
$u/d$-quark mass decreases from left panels to right panels.}
\label{fig:alleffmass}
\end{center}
\end{figure}

Figure~\ref{fig:alleffmass} shows effective masses of the $\eta$
and $\eta^\prime$ for all quark mass combinations. 
Because effective masses of the $\eta$ show approximate plateaus 
around $t\approx 4$, $m_\eta$ are determined from fits in 
the range $t=3-5$. 
For $\eta^\prime$, plateaus are clear for the lightest $u/d$-quarks,
while they are not so apparent for heavier $u/d$-quarks. 
Nonetheless off-diagonal parts of the diagonalized
propagator are consistent with zero already at $t=2$ for all cases,
which suggests that contribution from higher excited states is small.
Therefore, we estimate $m_{\eta^\prime}$ from fits in the range $t=2-4$.
Table~\ref{tab:results} summarizes masses of the $\eta$ and $\eta^\prime$.

For degenerate $u/d$ and $s$ quarks, the $\eta$ is a pure octet state, 
and therefore $m_\eta$ should agree with $m_{\eta_{ss}}$. 
Values of $m_\eta$ in Table~\ref{tab:results} are determined by 
the procedure explained above even for these cases. 
On the other hand, one-mass fits to $\eta_{ss}$ propagators 
in the range $t=8-16$ give $m_{\eta_{ss}}=$ 0.6251(13) and 0.5136(09) 
for $K_{ud}=K_s=$ 0.13710 and 0.13760 respectively.
They are larger than $m_\eta$ by about 2 percent.
We regard the 2 percent as an estimate of systematic error in $m_\eta$
arising from excited state contaminations.

\begin{table}
\begin{center}
\begin{tabular}{c|cc|cc}
 \hline
          & \multicolumn{2}{c|}{$K_s=0.13710$} 
          & \multicolumn{2}{c}{$K_s=0.13760$} \\
 $K_{ud}$ & $m_\eta$ & $m_{\eta^\prime}$  & $m_\eta$ & $m_{\eta^\prime}$ \\
 \hline
 0.13655 & 0.6605(35) & 0.816(33) & 0.6022(39) & 0.767(30) \\ 
 0.13710 & 0.6116(31) & 0.728(31) & 0.5645(23) & 0.709(32) \\
 0.13760 & 0.5588(34) & 0.678(34) & 0.5055(22) & 0.662(31) \\
 0.13800 & 0.5122(32) & 0.740(42) & 0.4555(23) & 0.667(39) \\
 0.13825 & 0.4751(45) & 0.629(29) & 0.4220(37) & 0.541(26) \\ 
 \hline	 
\end{tabular}
\caption{Masses of the $\eta$ and $\eta^\prime$ mesons.}
\label{tab:results}
\end{center}
\end{table}

\subsection{Chiral extrapolations}

\begin{figure}
\begin{minipage}{77mm}
\begin{center}
\includegraphics[scale=0.80]{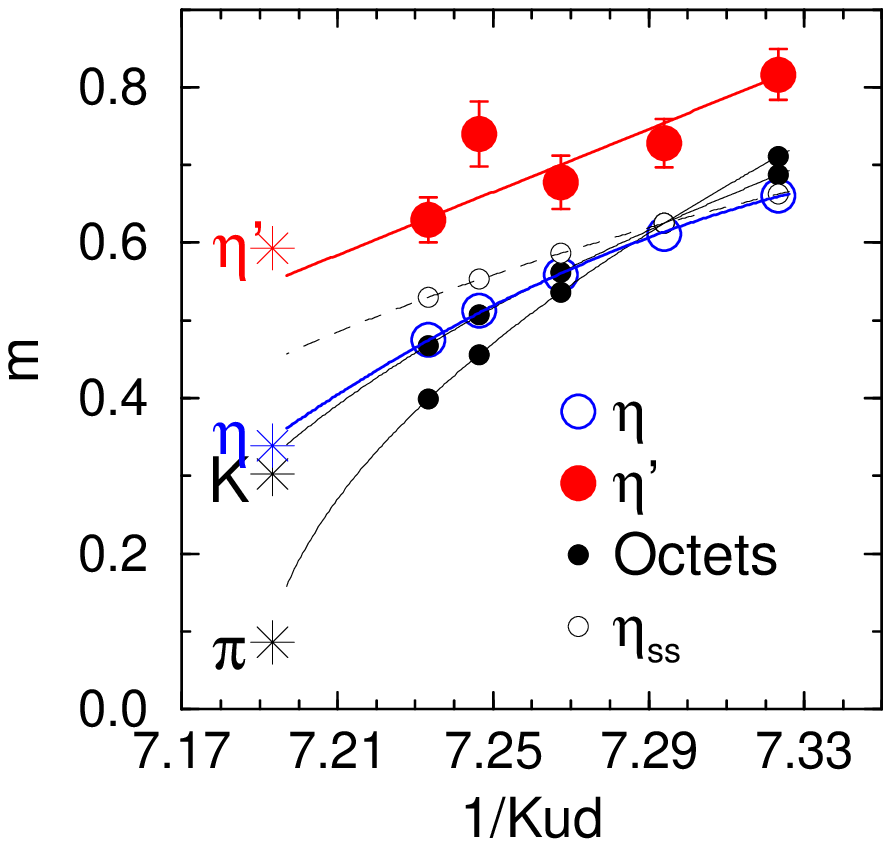}
\end{center}
\end{minipage}
\begin{minipage}{77mm}
\begin{center}
\includegraphics[scale=0.80]{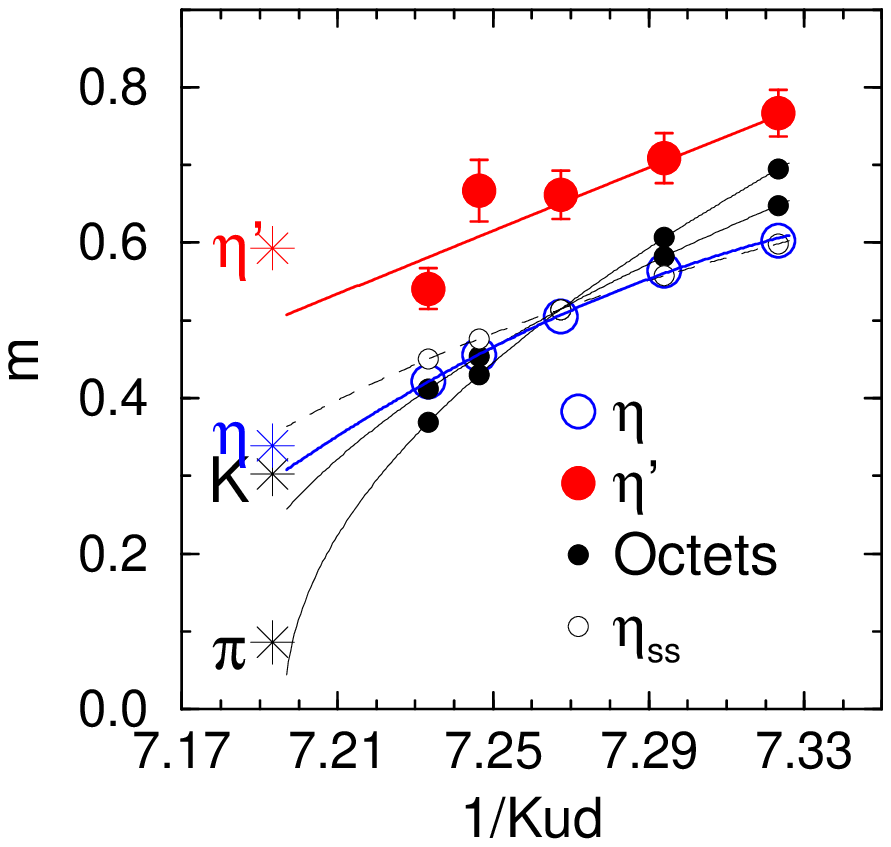}
\end{center}
\end{minipage}
\caption{Pseudoscalar meson masses versus $1/K_{ud}$ for 
$K_s$=0.13710 (left figure) and 0.13760 (right figure).
Stars marked at the physical $u/d$ quark mass point indicate 
experimental values.}
\label{fig:chiral}
\end{figure}

Chiral fits to $m_\eta$ and $m_{\eta^\prime}$ are made using low 
order polynomial functions of quark masses defined by
$m_{ud} = (1/K_{ud}-1/K_c)/2$ and $m_s=(1/K_s-1/K_c)/2$
($K_c$ is the critical hopping parameter).
Masses of pseudoscalar mesons are plotted in Fig.~\ref{fig:chiral}
as a function of $1/K_{ud}$. 
The figure suggests that $m_{\eta^\prime}$ shows a linear dependence
in $m_{ud}$ within our errors, so we assume a simple linear function
\begin{equation}
 m_{\eta^\prime} = A + B_{ud} m_{ud} + B_s m_s.
\end{equation}  
We obtain $\chi^2/\mbox{dof}$ = 1.4, which is acceptable.
We include a quadratic term of $m_{ud}$ for $m_\eta$, 
\begin{equation}
 m_{\eta} = C + D_{ud} m_{ud} + D_s m_s + E m_{ud}^2, 
\end{equation}  
because a linear function does not fit to data
due to curvatures shown in Fig.~\ref{fig:chiral}.
The quadratic term reduces $\chi^2$ of the fit greatly
to $\chi^2/\mbox{dof}$ = 1.0.
We note that linear fits to $m_{\eta,\eta^\prime}^2$
give masses at the physical point consistent with those from 
the fits above within $3 \sigma$ for the $\eta$ 
and $1 \sigma$ for the $\eta^\prime$.

Masses of the $\eta$ and $\eta^\prime$ extrapolated to 
the physical point, as determined from experimental 
$m_\pi$, $m_\rho$ and $m_K$, read
\begin{eqnarray}
 m_\eta          &=&   0.545(16) \ \mbox{GeV}, \\
 m_{\eta^\prime} &=&   0.871(46) \ \mbox{GeV}.
\end{eqnarray}

\section{Conclusions and Future Plan}\label{sec:Conclusion}
We have estimated the masses of the $\eta$ and $\eta^\prime$ in 
$N_f=2+1$ lattice QCD, albeit at one lattice spacing.
Our result $m_\eta$=0.545(16) GeV is consistent with 
the experimental value of 0.550 GeV.  
The $\eta^\prime$ meson mass $m_{\eta^\prime}$=0.871(46) GeV 
is also close to experiment, 0.960 GeV.  
This is encouraging since the small difference
of 100 MeV between our value for $m_{\eta^\prime}$ and experiment 
may well be accounted for by systematic errors.

Finite lattice spacing is one of possible origins of systematic
errors. Another shortcoming of this work is a relatively large 
value of $u/d$ quark masses. 
We expect to improve our program for the U(1) problem using
CP-PACS/JLQCD configurations already generated for finer lattices
and configurations which will be generated by the PACS-CS
collaboration~\cite{ref:pacscs} soon for lighter quarks. \\[0.2cm]
This work is supported by 
the Epoch Making Simulation Projects of Earth Simulator Center,
the Large Scale Simulation Program No.06-13 (FY2006) of 
High Energy Accelerator Research Organization (KEK),
the Large Scale Simulation Projects of 
Academic Computing and Communications Center, University of Tsukuba,
Inter University Services of Super Computers of 
Information Technology Center, University of Tokyo,
Super Sinet Projects of National Institute of Informatics,
and also by the Grant-in-Aid of the Ministry of Education
(Nos. 
13135204, 
13135216, 
15540251, 
16540228, 
16740147, 
17340066, 
17540259, 
18104005, 
18540250, 
18740130, 
18740167  
).

\end{document}